# The Evolution of Nova Remnants

Michael F. Bode

*Astrophysics Research Institute, Liverpool John Moores University,
Twelve Quays House, Birkenhead, Wirral, CH41 1LD, UK*

**Abstract.** In this review I concentrate on describing the physical characteristics and evolution of the nebular remnants of classical novae. I also refer as appropriate to the relationship between the central binary and the ejected nebula, particularly in terms of remnant shaping. Evidence for remnant structure in the spectra of unresolved novae is reviewed before moving on to discuss resolved remnants in the radio and optical domains. As cited in the published literature, a total of 5 remnants have now been resolved in the radio and 44 in the optical. This represents a significant increase since the time of the last conference. We have also made great strides in understanding the relationship of remnant shape to the evolution of the outburst and the properties of the central binary. The results of various models are presented. Finally, I briefly describe new results relating to the idiosyncratic remnant of GK Per (1901) which help to explain the apparently unique nature of the evolving ejecta before concluding with a discussion of outstanding problems and prospects for future work.

## INTRODUCTION

The observation and modelling of nova remnants are important from several standpoints. For example, imaging and spectroscopy of optically resolved remnants allow us to apply the expansion parallax method of distance determination with greater certainty than any other technique (it should be noted that without knowledge of the three-dimensional shape and inclination of a remnant, this method is prone to significant error). On accurate distances hang most other important physical parameters, including energetics and ejected mass. In addition, remnant morphology (and potentially the distribution of abundances) can give vital clues to the orientation and other parameters of the central binary and the progress of the TNR on the white dwarf surface. Finally, a fuller understanding of nova remnants has implications for models of the shaping of planetary nebulae and (in at least one case) physical processes in supernova remnants.

It is now of course almost thirteen years since the last major conference on Classical Novae, held in Madrid. At that time, a comprehensive review of the optical imagery of novae was given by Richard Wade [1] and of radio remnants by Bob Hjellming [2]. In the abstract of Wade's review, it was stated:

"There is room for much additional work in discovering new remnants and in characterising those that are known", and

"The mechanism that shapes the remnants is not yet known with certainty".

On both these counts, substantial progress has been made, particularly for the optically-resolved remnants.

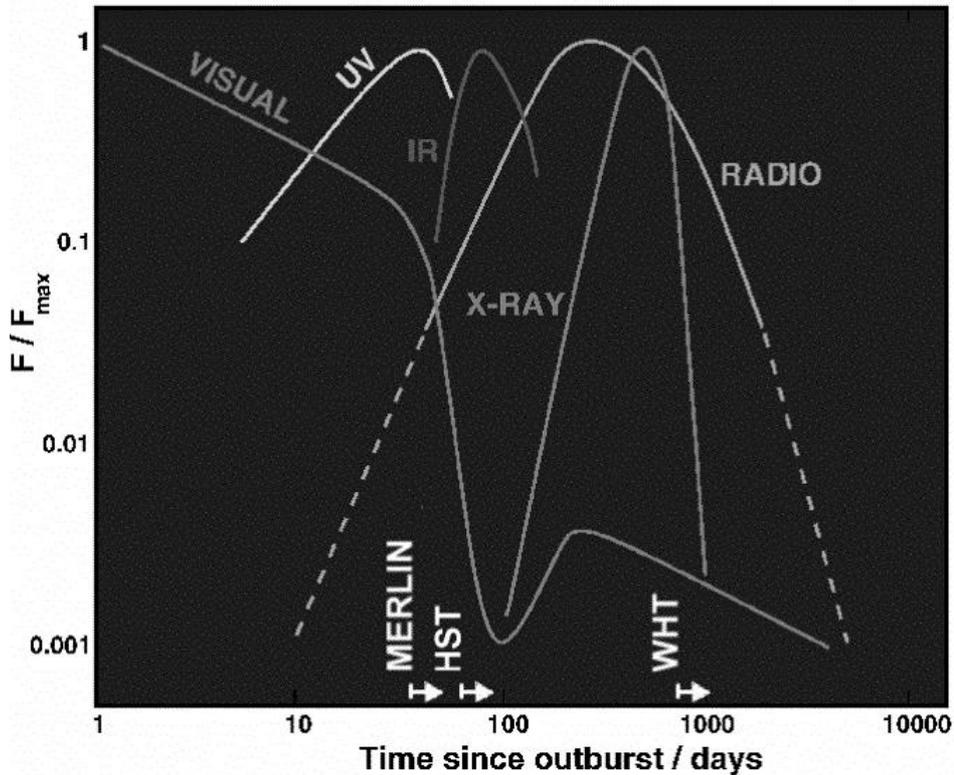

**FIGURE 1.** Schematic multi-frequency development of a nova outburst with times at which a remnant with $v_{exp}$= 1000 km s$^{-1}$, $d = 1$ kpc becomes spatially resolved in the radio (MERLIN), plus optically from space (HST) and on a conventional ground-based telescope (WHT).

# EVIDENCE FOR STRUCTURE IN UNRESOLVED REMNANTS

Intermediate to high-resolution spectroscopy of the spatially unresolved remnants of many novae has long been known to show evidence of organised structure (see e.g. [3] and review by Shore, this volume). This has allowed simple models of remnant geometry to be formulated and the existence of structures such as equatorial and tropical rings and polar caps/blobs to be inferred (see Fig. 2). However, optical depth effects (for the permitted lines) and variations in ionisation/excitation mean such analysis should be treated with caution. It is only when we spatially resolve the shell that we can (with care) become more confident in our conclusions.

In the case of HR Del (1967), Solf [4] performed heroic ground-based spectroscopic observations in exceptional seeing which led him to a more accurate appreciation of the true structure of the remnant (see paper by O'Brien, this volume, for recent optical imagery and spectroscopy of the resolved remnant which yield definitive results).

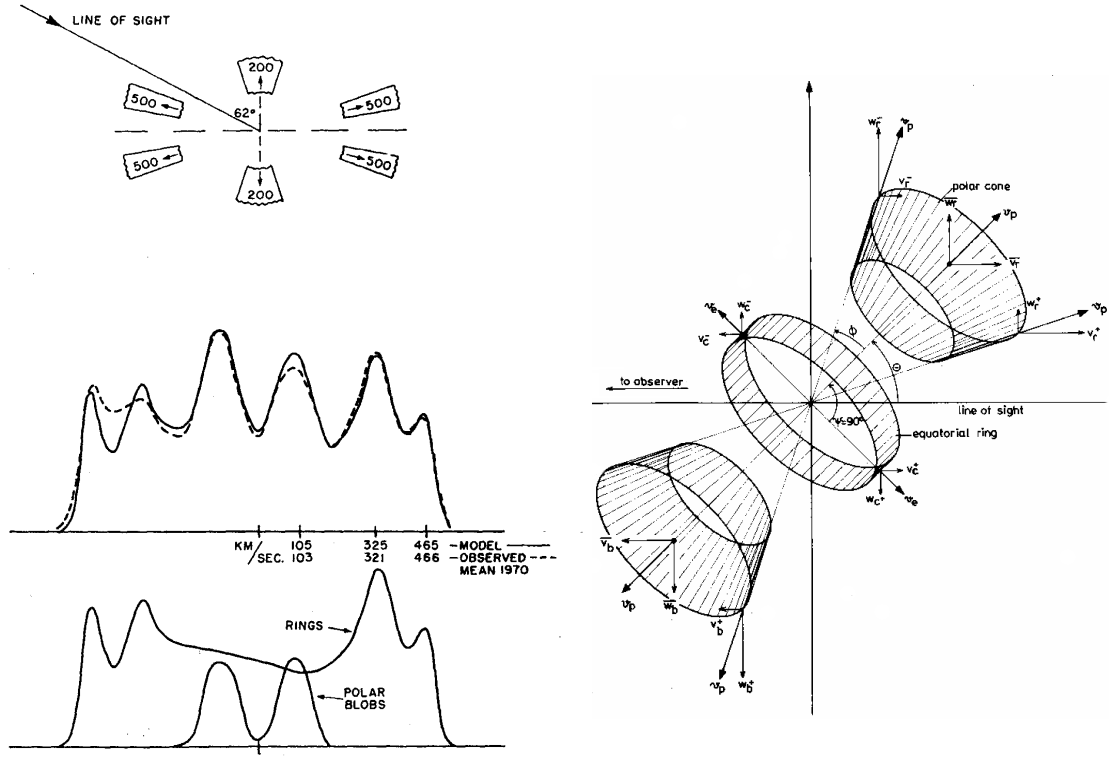

**FIGURE 2.** Mean Balmer line profiles in 1970 and computed model (left) for HR Del (reprinted from [3]) and (right) model of the same ejecta from high spatial resolution spectroscopy 14-15 years post-outburst (reprinted from [4]). Solf concluded that the remnant was a prolate ellipsoid with polar caps and equatorial ring, with an inclination $i = 38°$.

## EVOLUTION OF REMNANTS IN THE RADIO

Gehrz (this volume) gives a comprehensive review of the physical parameters that might be derived from a combination of infrared and radio observations of novae. It is evident from his review, and with reference to Fig. 1, that a typical nova can theoretically be resolved by a radio array such as MERLIN whilst still on the optically thick, rising part of the radio light curve. According to a simple model of radio evolution by Seaquist [5], the peak flux, angular size at peak and time to peak are given approximately by:

$$f_{max} \sim 43 \, (\nu / 5 \text{ GHz})^{1.16} \, (T_e / 10^4 \text{K})^{0.46} \, (M_{ej} / 10^{-4} M_\odot)^{0.8} \, (d / \text{kpc})^{-2} \text{ mJy}$$

$$\theta_{max} \sim 0.6 \, (\nu / 5 \text{ GHz})^{-0.42} \, (T_e / 10^4 \text{K})^{-0.27} \, (M_{ej} / 10^{-4} M_\odot)^{0.4} \, (d / \text{kpc})^{-1} \text{ arcsec}$$

$$t_{max} \sim 1.3 \, (\nu / 5 \text{ GHz})^{-0.42} \, (T_e / 10^4 \text{K})^{-0.27} \, (M_{ej} / 10^{-4} M_\odot)^{0.4} \, (v_{exp} / 1000 \text{ km s}^{-1})^{-1} \text{ yr}$$

Thus typically, ~1 yr after outburst, at the time of maximum in the radio light curve, the surface brightness at 5 GHz, $\Sigma_5 \sim 300$ µJy/beam with MERLIN (cf. ~ 50 µJy/beam rms noise) and the expanding remnant should be detectable at least until this time. Thereafter, $f_\nu \propto t^{-3}$ and $\Sigma_5 \propto t^{-5}$; i.e. unless the ejecta are clumped, or (less likely) $T_e$ is increasing, the remnant will rapidly become undetectable.

At the time of the last conference, Hjellming [2] listed only one nova (QU Vul) with a radio-resolved remnant (we do not include GK Per here for reasons we address below). Table 1 shows that in the published literature there are now four novae where the early evolution of the resolved remnant has been followed in some detail. We suspect however that there are more observations of novae in the VLA archive that deserve attention. It should also be noted that the extensive radio observations conducted by the late Bob Hjellming of V1974 Cyg (1992) have so far only been published in conference proceedings [6].

**TABLE 1. Radio-resolved Remnants**

| | Speed Class [7] | Time after outburst first reported resolved (days) | Primary Reference |
|---|---|---|---|
| QU Vul (1984) | F | 289 | [8] |
| V1974 Cyg (1992) | F | 80 | [9] |
| V705 Cas (1993) | MF | 585 | [10] |
| V723 Cas (1995) | VS | 840 | Heywood (this volume) |

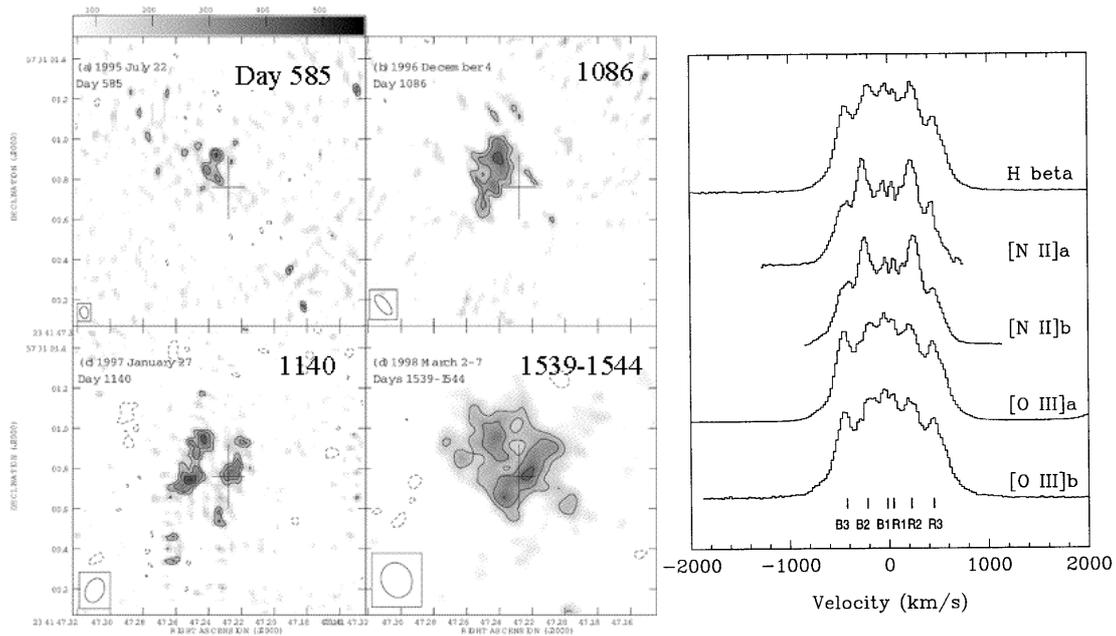

**FIGURE 3.** MERLIN maps at 5GHz showing the radio evolution of the remnant of V705 Cas (1993 - left, each box ~1.4″ on a side) and WHT optical spectra (right) taken on day 963 (reprinted from [10]).

Hjellming attempted to formulate a unified model of the radio and optical development of V1974 Cyg [6]. This involved a hybrid of the variable wind and Hubble flow models comprising a terminating wind with a linear velocity gradient (see review by Gehrz, this volume). In addition, he required the inner and outer shell boundaries to have different ellipsoidal shapes and an initial temperature rise was included. From this, he derived reasonable fits to the radio behaviour, but the fits to the spectral line shapes from HST observations were obviously too simplistic. Overall, this was a valiant attempt, but the model was largely phenomenological.

Figure 3 shows MERLIN maps of the complex evolution of the remnant of V705 Cas through the optically thick to early optically thin phases [10]. Also shown are WHT spectra from day 963 which imply an ordered structure (consistent with expanding equatorial and tropical rings in a remnant with $i = 60°$ – see also O'Brien, this volume). The optical spectra are difficult to reconcile with the radio observations. Indeed, the radio structure often seems to show drastic changes which make straightforward interpretation very difficult.

However, there are several fundamental complications with radio interferometric observations. First of all, larger-scale, lower surface brightness emission may be "resolved out". More importantly, it is impossible to disentangle the effects of changing optical depth and temperatures without high spatial resolution, simultaneous multi-frequency observations of the resolved shell. Currently with MERLIN for example, this is not possible. We will however address important future developments in the concluding section.

## OPTICAL IMAGING AND SPECTROSCOPY

Observations of resolved remnants in the optical are still the most fruitful as far as determining their physical characteristics and gaining insight into shaping mechanisms are concerned. Taking the typical nova expansion velocity and distance (Fig. 1), we would expect to first resolve the remnant after ~ 2 months with HST and ~ 2 years from the ground (without the aid of Adaptive Optics - see below).

**TABLE 2. Remnants Resolved Optically Since Wade (1990) Review [1]**

|  | Ground or Space | $t_3$ (days) | Ref. |
| --- | --- | --- | --- |
| DY Pup (1902) | GB | 160 | [11] |
| V450 Cyg (1942) | GB | 108 | [12] |
| CT Ser (1948) | GB | >100? | [13] |
| RR Cha (1953) | GB | 60 | [11] |
| HS Pup (1963) | GB | 65 | [11] |
| QZ Aur (1964) | GB | 23-30 | Esenoglu (this volume) |
| V3888 Sgr (1974) | GB | 10 | [13] |
| NQ Vul (1976) | GB | 65 | [12] |
| PW Vul (1984) | GB | 126 | [14] |
| QU Vul (1984) | GB | 49 | [15] |
| V1819 Cyg (1986) | HST | 89 | [13] |
| V842 Cen (1986) | GB | 48 | [11] |
| QV Vul (1987) | HST | 53 | [13] |
| V351 Pup (1991) | HST | 26 | Ringwald (this volume) |
| V1974 Cyg (1992) | HST | 42 | [16] |
| HY Lup (1993) | HST | >25 | [13] |
| V1425 Aql (1995) | HST | ~30 | Ringwald (this volume) |
| CP Cru (1996) | HST | 4 ($t_2$) | Ringwald (this volume) |

Table 2 gives details of all the optically-resolved nova remnants to appear in published literature since the Wade review [1] which cited 26. The total has substantially increased (a) because older remnants are inevitably becoming resolvable from the ground and (b) since 1989 we have seen the advent of HST.

## Ground-based Optical Imagery

The most extensive recent ground-based imaging surveys have been those of Slavin et al [12] for the northern hemisphere and Gill and O'Brien [11] for southern objects. Figure 4 shows a selection of results from these surveys.

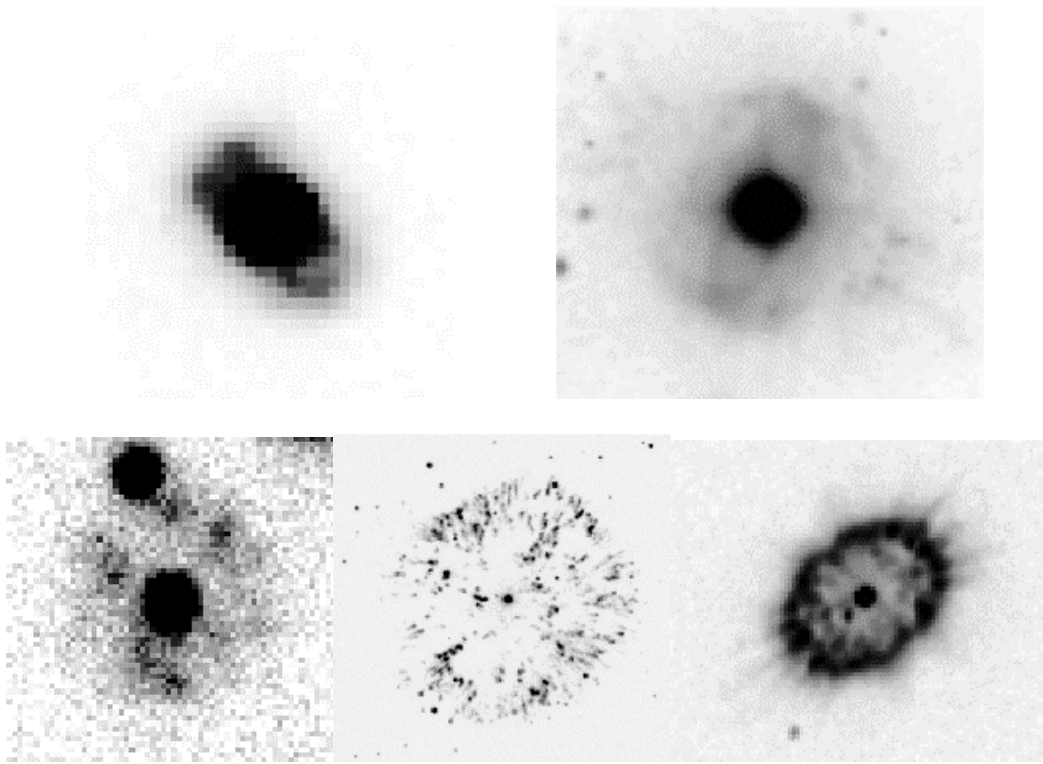

**FIGURE 4.** Ground-based optical images of novae of varying speed class (from Ref's [11] and [12]). Clockwise from top left: HR Del in [OIII] (outburst 1967, Very Slow, major axis ~12″); RR Pic in Hα/[NII] (1925, S, diameter ~ 21″); DQ Her in Hα (1934, MF, ~23″x17″ with halo ~ 47″x29″ and "tails" extending ~20″ from points of origin); GK Per in [NII] (1901, VF, ~90″ diameter), and V1500 Cyg in Hα (1975, VF, diameter ~8″).

The optical images tend to confirm the presence of structure as implied from earlier spectroscopy. For example, in RR Pic and DQ Her, equatorial (and for DQ Her, tropical) rings are apparent. In addition, in both these objects, "tails" of emission are evident, streaming away from knots in the shell, suggestive of ablation by faster-moving ejected material (this process is discussed further by O'Brien, this volume). There also appears to be systematically *less* shaping with *increasing* speed class ([12], [13] - see below). Indeed, as pointed out in [12], the very fast nova GK Per, if placed at the distance of V1500 Cyg, would show a very similar, relatively circular, but otherwise amorphous remnant.

# HST Imaging and Ground-based Optical Spectroscopy

The combination of HST imaging, long-slit optical spectroscopy and simple modelling has been shown to be a very powerful tool in determining the precise geometry of resolved remnants. It is also potentially important in exploring other physical properties, such as possible abundance gradients. The existence of these would have significant consequences for our understanding of the progress of the TNR across the WD surface. Previous work (e.g. [17]) has not been conclusive in this regard. Enhanced emission in [NII] from equatorial rings (such as that seen for example in Fig. 5, from the work of Gill and O'Brien [18] [19]) may not be due to a simple overabundance of nitrogen. Further discussion on this point is given in O'Brien (this volume), but there is clearly much more work to be done in this regard.

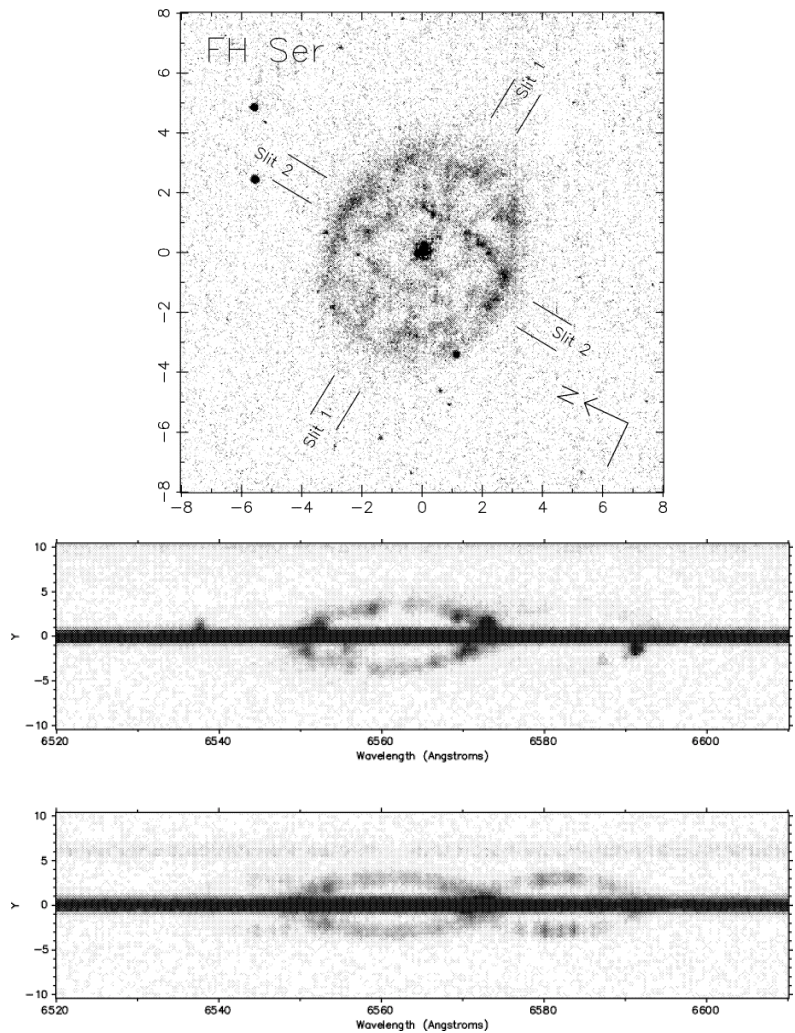

**FIGURE 5.** HST observations of FH Ser (1970 - top) through the F656N filter (axes in arcsec), taken from [19]. Combined with WHT long-slit spectroscopy (bottom, slit 1 top panel) and modelling, this confirmed that the nova shell is prolate, axial ratio = $1.3 \pm 0.1$, $i = 62° \pm 4°$, equatorial expansion velocity = $490 \pm 20$ km s$^{-1}$ and hence $d = 950 \pm 50$ pc [19]. The clear [NII] ring most likely delineates the plane of the orbit of the central binary.

HST results provide further confirmation of a relationship between the degree of shaping and speed class. In Fig. 6 we have included corrections for inclination where possible. This correction appears to strengthen the correlation further.

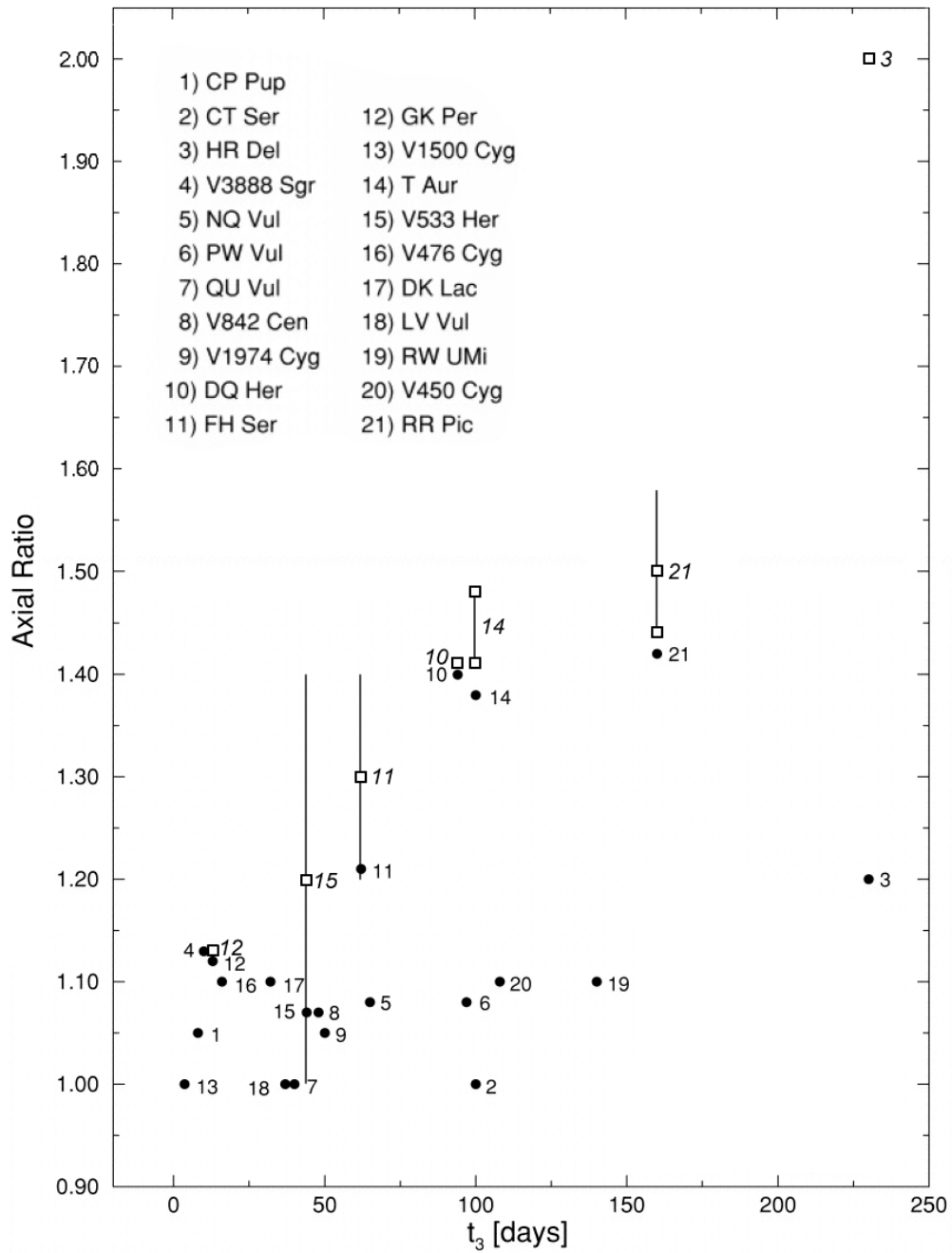

**FIGURE 6.** Axial ratio of remnants versus speed class. Filled circles from [13]. Open squares show inclination-corrected results (data from [12], [18], [19], plus O'Brien, this volume, for HR Del).

# HYDRODYNAMICAL MODELLING

It is easily shown that the ejecta from the WD surface (and indeed the consequent pseudo-photospheric radius) rapidly envelop the secondary star yielding effectively a common-envelope binary. Early results of modelling the effects of the frictional deposition of energy and angular momentum by the secondary on the ejected nova envelope were reported at the Madrid meeting by Shankar et al. [20], [21]. Their 2-D hydrostatic wind models were essentially applicable only to restricted cases of very slow novae.

Subsequently, Lloyd et al. [22] used a 2.5-D hydro-code to investigate remnant shaping for a variety of speed classes. An important ingredient is the observed relation between ejection velocities and speed class [7]. In the case of slower novae for example, lower ejection velocities would lead to longer effective interaction times between the secondary and the ejecta, and hence more shaping might be expected. The basic model involves ejecta in the form of a wind with secularly increasing velocity and decreasing mass-loss rate. This evolved from a model of the early X-ray emission from V838 Her (1991) involving the interaction of ejecta with different expansion velocities early in the outburst [23]. The Lloyd et al. model produces rings, blobs and caps, plus a correlation of speed class to axial ratio in the sense required. However, it also produces *oblate* remnants.

Porter et al. [24] modified this basic model to include the effects of the rotating accreted envelope on the surface of the WD. From consideration of the effective gravity due to envelope rotation, and its effects on local luminous flux driving mass loss at outburst, a mass-loss rate and terminal velocity of ejecta were derived that are dependent on latitude on the WD. Figure 7 shows the comparison of results for a moderately fast nova without and with rotation of the accreted envelope. The Porter et al. models produce prolate shells as required. It should also be noted that a rather more satisfactory fit to the early radio evolution of V1974 Cyg may be provided by such models [25].

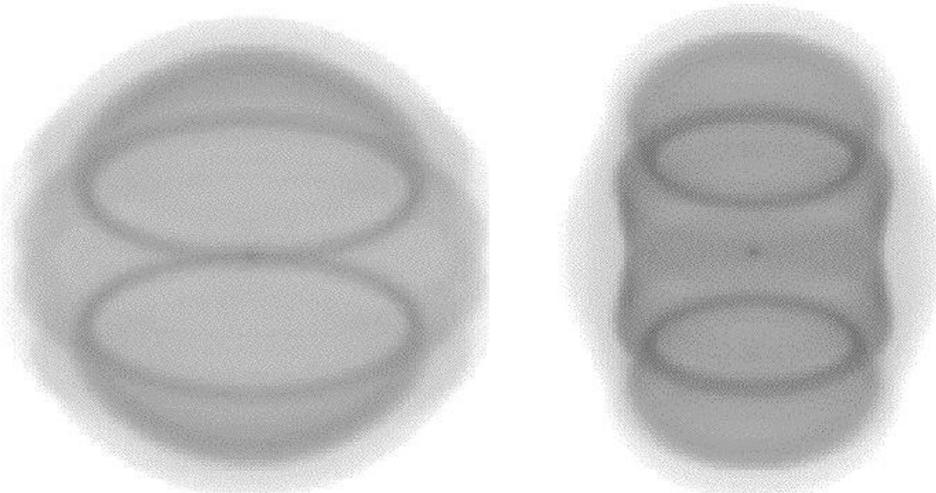

**FIGURE 7.** Synthetic images of remnants produced for the parameters of the ejecta and central binary of a moderately fast nova. Left - no envelope rotation. Right - accreted envelope rotating at 0.7 of the Keplerian velocity at the WD surface (taken from [24]).

# GK PER (1901) - A NOVA SUPER-REMNANT

As reported by Balman in these proceedings, GK Per is an unusual remnant in many respects (see also [26] and references therein). It is for example the only unequivocally non-thermal radio remnant so far discovered among the CNe This, together with the presence of extended X-ray emission and deceleration of the ejecta, has led to the conclusion that the high-velocity material from the 1901 outburst is interacting with a pre-existing circumstellar medium. As shown in Fig. 8, there is evidence in the far-infrared and optical for extended nebulosity that may be associated with a previous phase of the evolution of the central binary (most likely a born-again AGB star [27]).

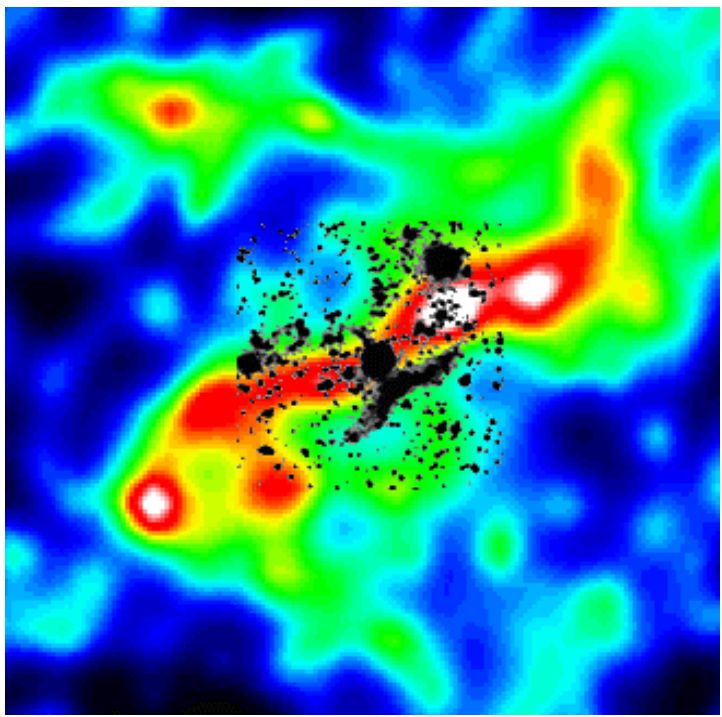

**FIGURE 8.** 100µm map from IRAS observations of the environs of GK Per [27], superimposed on which is a map of Hα emission [28]. The Hα image is 21′ on a side. The remnant of the 1901 nova explosion lies at the centre of the IRAS emission near the cusp of the Hα nebula and is now ~1′ in diameter (N at top, W to right). Note that the main interaction region in the nova remnant is to the SW.

One of the major questions has however been why the remnant is so asymmetric with the interaction largely being confined to the SW quadrant. A combination of Isaac Newton Telescope Wide Field Camera imagery and measurement of proper motion from archival plates over a 90-year baseline has confirmed our suspicion that the *whole system* is moving towards the SW. The large-scale "planetary" from the previous evolutionary phase is slowed by its interaction with the ISM. The nova binary travels through this essentially as a bullet. Then in 1901 we saw the very first outburst of the nova. The high velocity nova ejecta naturally impact a higher density of material in the direction of the whole system's motion than elsewhere [29]. Thus, much of the mystery is solved and the important place of GK Per in our understanding of CNe evolution is reinforced. In addition, this remnant, evolving as it does on a time scale two or three orders of magnitude shorter than that of a supernova remnant, and being close enough to resolve on relatively small length scales, is worthy of study in terms of furthering our understanding of the general physical processes occurring in such circumstances.

## CONCLUDING REMARKS

We have made major strides in understanding the evolution of nova remnants since the Madrid conference. However, as always, there are still outstanding challenges. These include: the reconciliation of remnant development in the optical and radio (and from which we would particularly aim to determine more accurately the all-important ejecta masses); full 3-D treatment of the shaping of nova remnants and the exploration of a fuller parameter space of models; following the evolution of "clumpiness" and relating this to dust formation (see papers by Evans and Gehrz, this volume), and untangling excitation/ionisation effects from true abundance gradients in resolved remnants.

We are fortunate however to have exceptional observational tools at our disposal, either existing or planned, to tackle these problems head on. In the short term, HST plus STIS will be important in determining whether there really are abundance gradients in the extended nebulae which in turn might be related to variations in the initial conditions of the TNR on the WD surface. In the radio, both the e-MERLIN and EVLA projects will provide superb opportunities for using radio imaging to its full potential. e-MERLIN for example will have full frequency switching and around 30 times the sensitivity of the current array, coupled with resolution as high as 8 m.a.s. This gives the real promise of untangling the effects of temperature and optical depth changes which otherwise lead to misleading interpretations of the evolution of remnants in the radio. Finally, advances in ground-based telescopes, including adaptive optics (see Diaz, this volume) and optical interferometer arrays hold out the prospect of following the evolution of the ejecta from very early in the outburst. The proposed Large Optical Array for example would be able to resolve the remnant of a typical nova within hours of the outburst - a time at which the mechanisms of shaping might still be in full swing.

## ACKNOWLEDGMENTS

I would like to thank my many colleagues from whom I have learnt so much over the years. In particular Tim O'Brien, John Porter, Huw Lloyd, Stewart Eyres, Richard Davis and Nye Evans, plus Franz Kahn, who is sorely missed. Thanks also to Andy Newsam and David Hyder who gave me invaluable help in the preparation of the figures for the review. Permission to reproduce figures from previously published works (as cited as appropriate in the text) was gratefully received from Blackwell Publishing, Astronomy and Astrophysics, and the American Astronomical Society.